\documentclass[epj,twocolumn]{webofc}

\usepackage{graphicx}
\usepackage{microtype}
\usepackage[varg]{txfonts}

\wocname{epj}
\woctitle{Seismology of the Sun and the Distant Stars 2016}

\newcommand{\st}[1]{_\text{#1}}
\newcommand{\Mm}{\rm\thinspace Mm}
\newcommand{\Msun}{\hbox{$\rm\thinspace \text{M}_{\odot}$}}
\newcommand{\uHz}{\hbox{\rm\thinspace $\mu$Hz}}

\begin{document}

\title{Surface effects in solar-like oscillators}
\author{\firstname{Warrick~H.~}\lastname{Ball}\inst{1,2}
  \fnsep\thanks{\email{wball@astro.physik.uni-goettingen.de}}}

\institute{
  Institut f\"ur Astrophysik, Georg-August-Universit\"at G\"ottingen, 
  Friedrich-Hund-Platz 1, 37077 G\"ottingen, Germany\\
  \email{wball@astro.physik.uni-goettingen.de}\and 
  Max-Planck-Institut f\"ur Sonnensystemforschung, 
  Justus-von-Liebig-Weg 3, 37077 G\"ottingen, Germany}

\abstract{Inaccurate modelling of the near-surface layers of solar
  models causes a systematic difference between modelled and observed
  solar mode frequencies. This difference---known as the ``surface
  effect'' or ``surface term''---presumably also exists in other
  solar-like oscillators and must somehow be corrected to accurately
  relate mode frequencies to stellar model parameters. After briefly
  describing the various potential causes of surface effects, I will
  review recent progress along two different lines. First, various
  methods have been proposed for removing the surface effect from the
  mode frequencies and thereby fitting stellar models without the
  disproportionate influence of the inaccurate near-surface
  layers. Second, three-dimensional radiation hydrodynamics
  simulations are now being used to replace the near-surface layers of
  stellar models across a range of spectral types, leading to
  predictions of how some components of the surface effect vary
  between stars. Finally, I shall briefly discuss the future of the
  problem in terms of both modelling and observation.}

\maketitle

\section{Introduction}
\label{s:intro}

The era of space-based asteroseismology, driven chiefly by COROT
\citep{corot} and \emph{Kepler} \citep{kepler}, has provided
observations of hundreds of cool main-sequence stars in which dozens
of individual mode frequencies can be measured.  To exploit this data,
however, we need to correct for a systematic difference between
observed and modelled mode frequencies caused by improper modelling of
the near-surface layers of these stars: the so-called \emph{surface
  term} or \emph{surface effect}.  Motivated by a newfound need to
correct for the surface effect, significant progress has been achieved
in the last few years and can be expected in the near future.

The purpose of this review is to first briefly recount our physical
understanding of the surface effect (Sec.~\ref{s:problem}) and then
review recent progress along two lines.  First, several authors have
proposed parametrizations of the surface effect (as a function of
frequency) to suppress its influence when fitting stellar models to
observed mode frequencies (Sec.~\ref{s:param}).  Second, a few
research groups have begun replacing the near-surface layers of
stellar models with average structures taken from detailed
three-dimensional radiation hydrodynamic simulations (3D RHD,
Sec.~\ref{s:3drhd}).  Finally, I close with a few thoughts on how we
might progress further on the problem of surface effects in the near
future (Sec.~\ref{s:future}).

I do not pretend that this review is exhaustive.  Judging by the
amount of material I excluded from my talk, it would be impossible to
cover all the literature on the subject in 30 minutes.  I apologize to
anyone who feels their contribution has been omitted and seek to
assure them that the cause is only brevity, not malice!

\begin{figure}
\centering
\includegraphics[width=\hsize,clip]{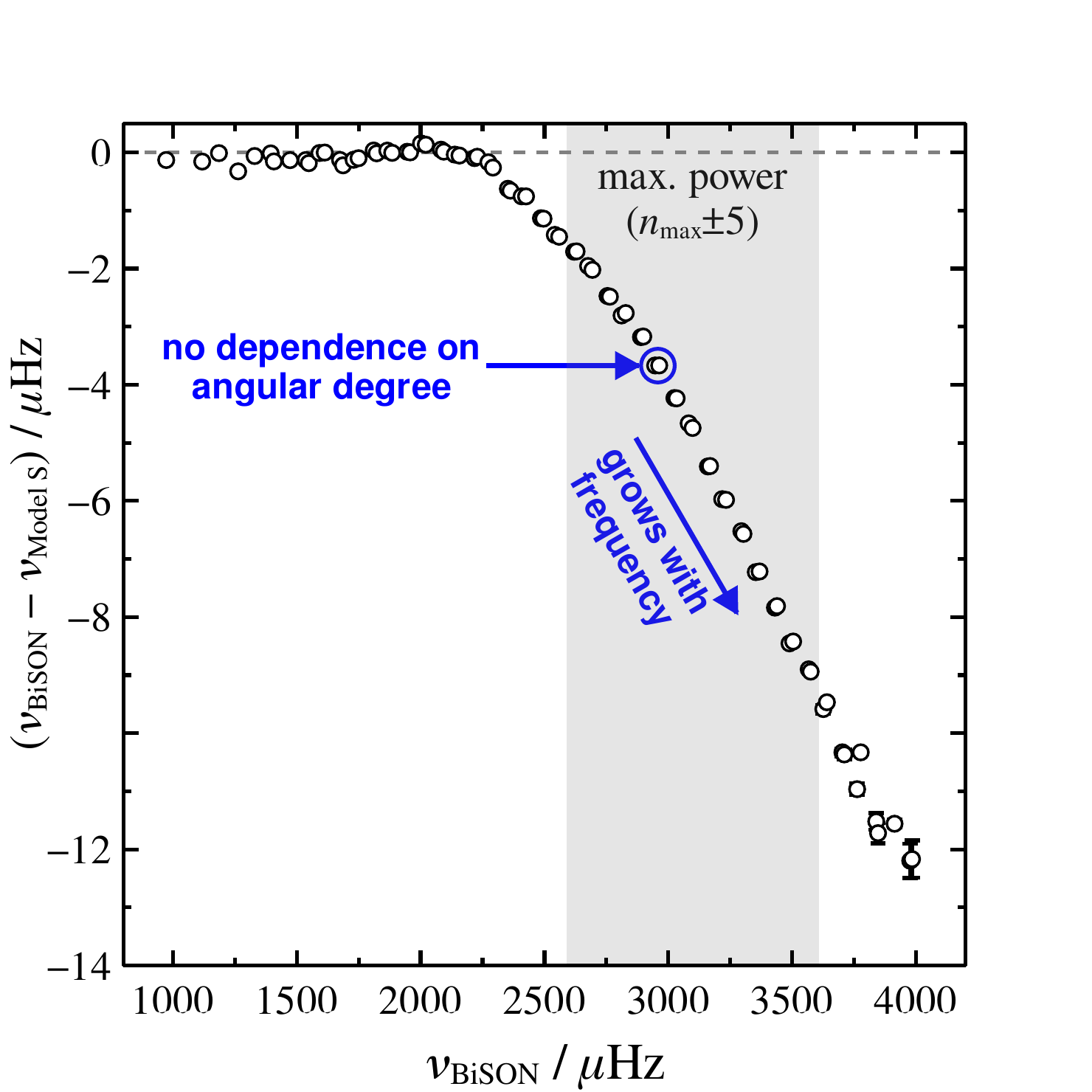}
\caption{The basic problem of surface effects in the Sun.  The white
  points show differences between low-degree ($\ell\leq3$) mode
  frequencies observed by the Birmingham Solar Oscillation Network
  \citep[BiSON; ][]{broomhall2009,davies2014a} and linear adiabatic
  mode frequencies predicted for a standard solar model \citep[Model
  S; ][]{modelS}.  The blue annotations show the two main features
  that lead us to believe that the systematic difference is a surface
  phenomenon: the differences grow with frequency and are largely
  independent of the angular degree $\ell$ (see text).  The shaded
  region shows where the modes have their greatest power and are most
  easily observed, demonstrating that for most asteroseismic targets,
  \emph{all} the observed frequencies are probably affected by the
  surface effect.}
\label{f:problem}
\end{figure}

\section{The problem}
\label{s:problem}

\subsection{Phenomenology of the surface effects}

Suppose that one calibrates a solar model in the traditional sense by
varying the mixing length parameter, initial helium abundance and
initial metallicity to evolve a stellar model that matches the Sun's
current radius, luminosity and surface metallicity, with the mass
fixed at $1\Msun$ and the age at the meteoritic age of the solar
system.  We take Model S \citep{modelS} as an example.  If we compute
the linear adiabatic mode frequencies of this model and compare them
with observed values \citep[e.g. the low-degree data from the
Birmingham Solar Oscillation Network, BiSON;][]{broomhall2009,
  davies2014a} we might hope that the differences between observed and
modelled frequencies are randomly scattered about zero.  

Instead, one gets the values plotted in Fig.~\ref{f:problem}.  The
white points indicate the differences between the mode frequencies
predicted for Model S and those observed by BiSON.  The discrepancy is
much larger than the quoted uncertainties but it is also not random,
and its structure tells us something about where the problem arises.
First, the frequency differences do not depend on the angular degree
$\ell$, which suggests that the discrepancy lies well above the modes'
lower turning points.  The low-degree data alone only tells us that
the cause is not very deep in the Sun but the frequency differences of
the higher-degree modes are also $\ell$-independent.  Because they
have shallower lower turning points, this suggests that the problem is
quite close to the Sun's surface.  Second, the frequency differences
are close to zero at frequencies below about $2200\uHz$.
Higher-frequency modes have shallower upper turning points, which
again implies that the problem is somewhere near the Sun's surface.
At $2200\uHz$, the modes' upper turning points are around $1\Mm$ below
the surface, which implies that the effect really is confined to the
near-surface layers.

The shaded region in Fig.~\ref{f:problem} indicates the range of
frequency covering about five radial orders either side of the
frequency of maximum oscillation power $\nu\st{max}$.  This is the
range in which modes oscillate with the greatest power and are thus
most easily observed.  It shows that the surface effect probably
affects nearly all the observed modes in distant Sun-like stars,
unlike the Sun, in which we observe low-frequency modes that appear
unaffected by the surface effect.  The best targets from the nominal
\textit{Kepler} mission have lowest mode frequencies equivalent to
about $2300\uHz$ in the Sun, which is within the range of affected
modes.  For this reason, the surface effect is unavoidable: when
fitting stellar models to individual mode frequencies, something
\emph{must} be done about the surface effect.  In the case of the Sun,
even the large separations of the modelled and observed frequencies
differ by about $1\uHz$.  When applied to the standard scaling
relations \citep{kjeldsen1995}, this bias in the large separation
corresponds to biases in mass and radius of about $3$ and $1.5$ per
cent, respectively.\footnote{The scaling relations are in essence
  empirical, which suppresses this effect.  But if the surface effect
  varies significantly between different stars, it could be
  important.}

\begin{figure*}
\centering
\begin{tabular}{r@{}l}
\includegraphics[width=0.5\hsize,clip]{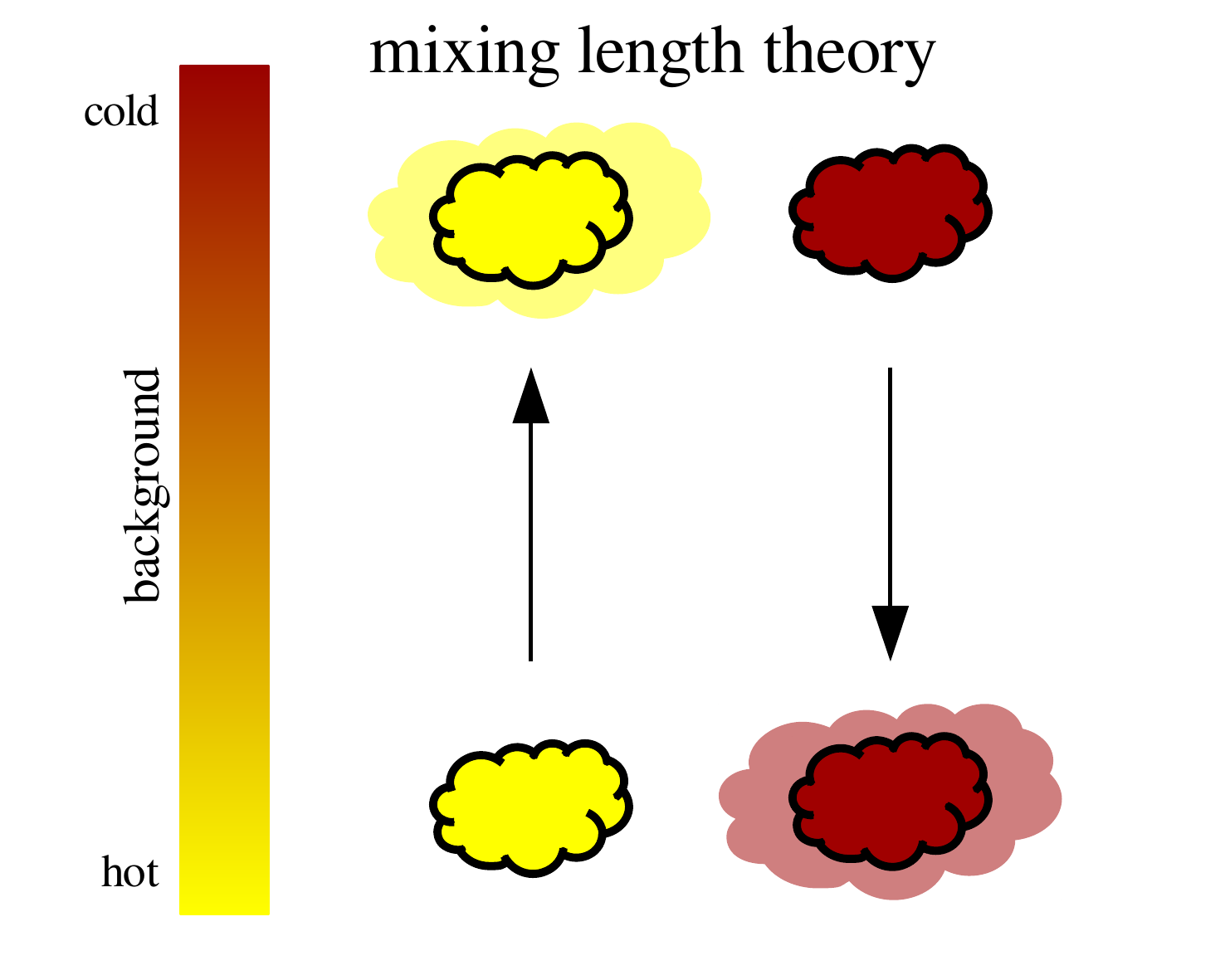} &
\includegraphics[width=0.5\hsize,clip]{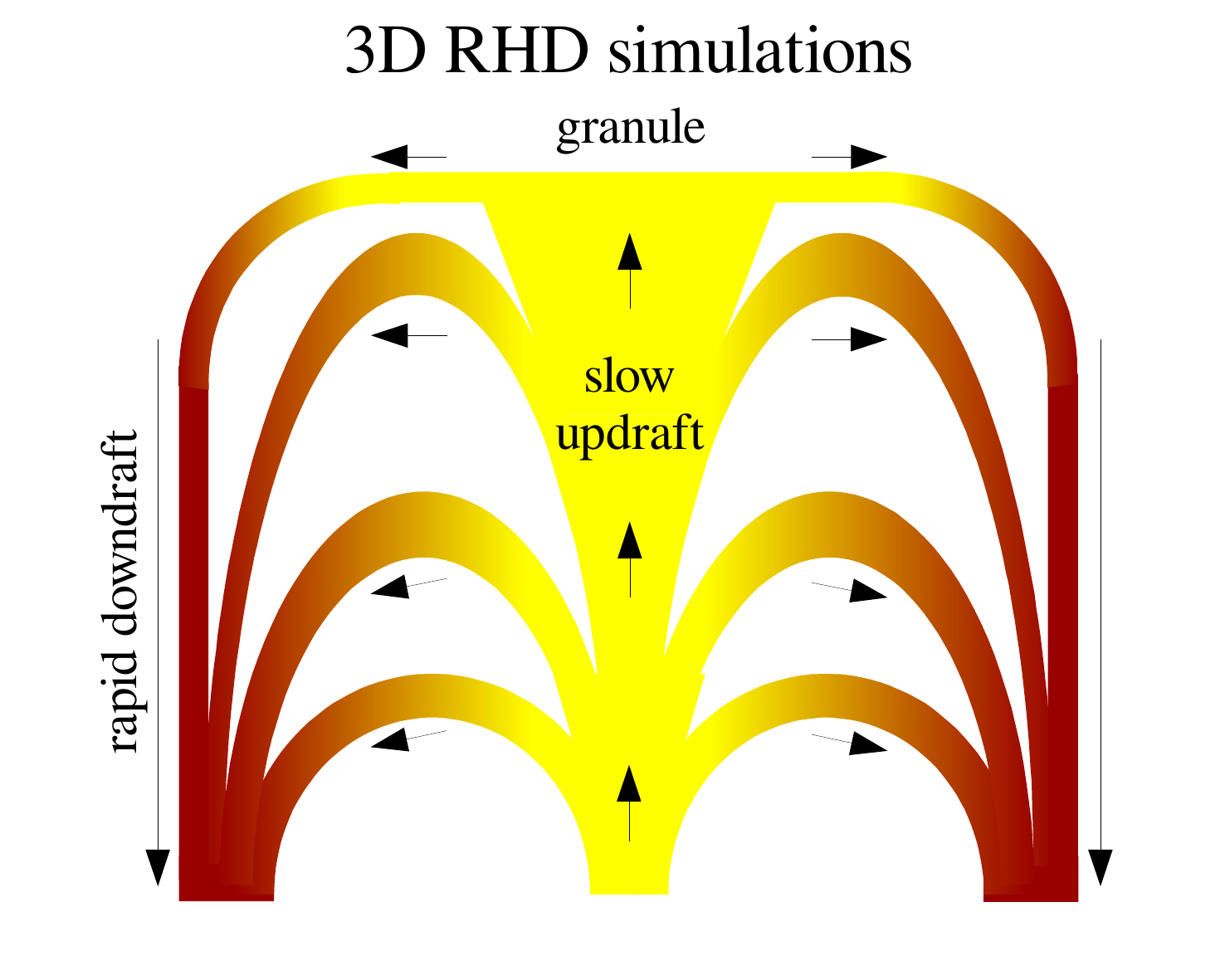}
\end{tabular}
\caption{A crude illustration of the difference between the
  mixing-length theory of convection (MLT, left) and the structure of
  near-surface convection suggested by 3D RHD simulations (right).  In
  MLT, buoyantly-unstable parcels of material retain their composition
  and heat content while floating upwards (or sinking downwards) by one
  \emph{mixing length} $l\st{MLT}$ before dispersing their composition
  and heat into their new surroundings.  In the 3D RHD simulations,
  slow, broad upflows expand as they rise through layers of decreasing
  density, ultimately reaching the surface and manifesting as
  granules.  At the surface, material cools and sinks back down
  between the granules in rapid, cool downdrafts that we see as
  intragranular lanes.}
\label{f:diagram}
\end{figure*}

\subsection{The physical cause}

It may come as a surprise that we have a fairly good idea about what
causes the surface effect: improper modelling of near-surface
convection.  Most stellar models use some form of mixing-length theory
(MLT), in which the convection zone is presumed to contain
buoyantly-unstable rising and falling parcels of material (see
Fig.~\ref{f:diagram}, left).  These parcels rise or fall by one mixing
length, typically parametrized in terms of the local pressure scale
height, $H_P=-\mathrm{d}r/\mathrm{d}\ln P$, after which they disperse,
mixing the heat and composition of their origin into their new
surroundings.

In reality, the flows are much more complicated, as is now understood
from detailed 3D RHD simulations that accurately reproduce many
observable features of convection at the Sun's surface \citep[see][
for an excellent review of the Sun's surface
convection]{nordlund2009}.  Let us start with one of the slow upflows.
As it rises and the density decreases, so the flow expands
horizontally and, to conserve mass, part of it must turn over and join
whatever downflows exist (see Fig.~\ref{f:diagram}, right).  The
rising plume ultimately appears as a granule at the surface, where the
flows are chiefly horizontal.  They then radiate heat to the vacuum of
space before plummeting downward in narrow, turbulent intragranular
lanes.  Along the way back down, these downflows will draw material
turning over from the widening upflows.

This is a very different picture from the calm rise and fall of MLT's
parcels and it leads to a number of effects that affect the mode
frequencies.  Following the thorough discussion by Rosenthal
\citep{rosenthal1997}, we can broadly divide these into two types of
effect.  \emph{Model} physics includes everything that is wrong with
the background model that we perturb.  This includes, but is not
limited to, MLT's incorrect temperature gradient, the incorrect
atmospheric structure and the absence of turbulent pressure.
\emph{Modal} physics includes everything that is wrong with the
calculation of the mode frequencies, which are affected by the
perturbation to the turbulent pressure \citep[e.g.][]{rosenthal1999},
the modification of wave speeds when travelling with or against the
flows \citep[e.g.][]{brown1984} and various effects of
non-adiabaticity \citep[e.g.][]{houdek1996}.  All of these effects are
most pronounced near the surface where convection becomes inefficient
and the temperature gradient deviates furthest from the adiabatic
value.

That so many physical effects contribute to the surface effect makes
it a difficult problem to tackle piece by piece.  In working on one
component, one might think the problem is solved, only to find that
another component returns you to square one.  But it is not hopeless!
We can learn how much each component might contribute and gradually
add them up, bearing in mind that as our models improve, we might
sometimes veer further from the observations before once again closing
the gap.

\begin{figure}
\centering
\includegraphics[width=\hsize,clip]{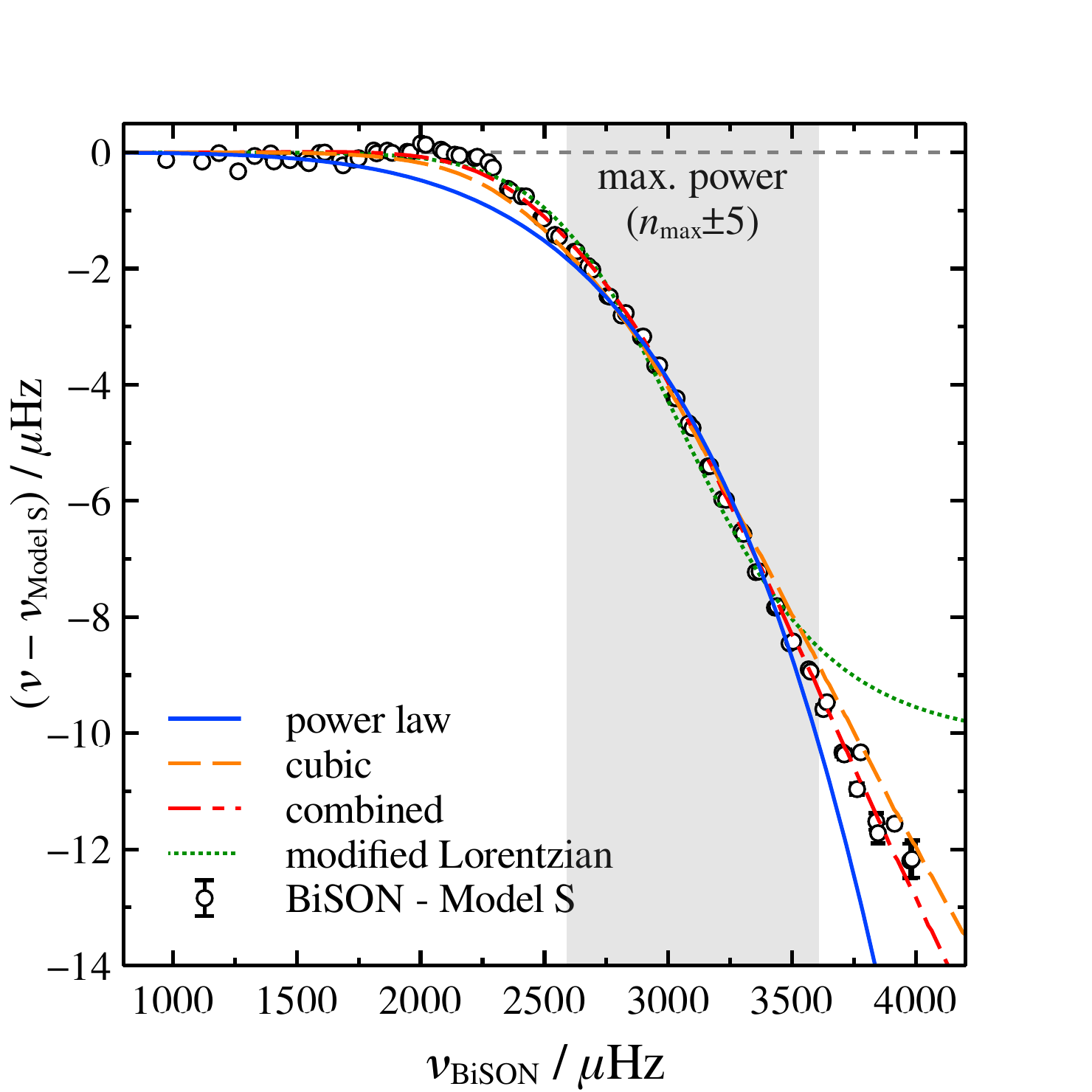}
\caption{Frequency differences, as a function of frequency, for the
  observed surface effect in the Sun (white points) and several
  parametrizations (see Sec.~\ref{s:param}).  The shaded region shows
  the frequency range over which the modes have their greatest power,
  as in Fig.~\ref{f:problem}.  The parametrizations shown are a
  power-law \citep[solid blue, ][]{kbcd2008}, the cubic and combined
  formulae \citep[dashed orange and dash-dotted red, ][]{ball2014} and
  a modified Lorentzian \citep[dotted green, ][]{sonoi2015}.}
\label{f:param}
\end{figure}

\section{Parametrizations}
\label{s:param}

The surface effect in Fig.~\ref{f:problem} appears to be a relatively
simple function of mode frequency only.  Thus, several groups have
proposed parametric forms for this function whose parameters can be
fit when comparing stellar models to observations.  Here I shall
review the best known and compare them for the Sun.

First, Kjeldsen et al.\,\citep{kbcd2008} proposed that the surface effect can be
described as a power law with an index fixed to a solar-calibrated
value (usually around $5$ but slightly dependent on the precise
physics of the stellar model).  They also proposed that the magnitude
of the power law be fit after rescaling the frequencies so that the
stellar model being compared has the same mean density as the observed
star.  To rescale the frequencies so, they propose using the ratio of
the large separations.  This simple parametrization has been widely
used since its publication \citep[e.g.][]{silva2013}.

More recently, Ball \& Gizon\,\citep{ball2014} proposed
parametrizations based on surface perturbations and the asymptotic
behaviour of the eigenmodes.  Roughly speaking, the displacement
eigenfunctions are exponentially decaying functions near the
photosphere and, combining them with the variational principle for the
linear, adiabatic oscillation equations \citep{ledoux1958}, one finds
that, for a sound speed perturbation or pressure scale height
perturbation near the surface, the frequency shifts go either like
$\nu^3/\mathcal{I}$ or $\nu^{-1}/\mathcal{I}$, where $\nu$ is the mode
frequency and $\mathcal{I}$ the normalized mode inertia.  These
parametric forms, which Ball \& Gizon\,\citep{ball2014} refer to as
the \emph{cubic} and \emph{inverse} terms, respectively, were
originally derived by Gough\,\citep{gough1990}\footnote{The cubic term
  is also mentioned by Libbrecht \& Woodard\,\citep{libbrecht1990} and
  Goldreich et al.\,\citep{goldreich1991}.} in a discussion of the
Sun's frequency shifts over the magnetic activity cycle.  The inverse
term alone does not fit the data well so Ball \&
Gizon\,\citep{ball2014} proposed to combine it with the cubic term,
giving what they call the \emph{combined} surface correction.

Most recently, Sonoi et al.\,\citep{sonoi2015} proposed to describe the surface
effect as a modified Lorentzian function and calibrated its parameters
to frequency shifts induced by replacing the near-surface layers of
stellar models with averaged data from hydrodynamics simulations (see
Sec.~\ref{s:3drhd}, below).  This parametrization is very new and has
not yet been tested on observed data.

Fig.~\ref{f:param} shows the same data as Fig.~\ref{f:problem} (BiSON
against Model S), along with the above-mentioned parametrizations.
The power law fit performs reasonably well in the shaded range around
$\nu\st{max}$ but overestimates the surface effect both where it
begins to rise and at higher frequencies.  The cubic and the combined
terms fare better.  Though the improvement by using the combined term
(rather than just the cubic term) is significant for the Sun, this was
not the case for the COROT target HD~52265 studied by Ball \&
Gizon\,\citep{ball2014}.  Finally, the modified Lorentzian captures
most of the low-frequency behaviour but underestimates the difference
at high frequencies.

Though different in principle, it is worth mentioning several methods
proposed by Roxburgh (and Vorontsov in earlier work)
\citep{roxburgh2003, roxburgh2015, roxburgh2016}.  These are all based
on representing the oscillation modes as simple oscillations with
phase shifts at the inner and outer boundaries.  The outer phase shift
contains the undesired and presumably $\ell$-independent surface term
whereas the inner phase shift is related to the structure of the
stellar core.  One can combine the frequencies into ratios of
differences or so-called \emph{separation ratios} that are nearly
independent of the near-surface layers \citep{roxburgh2003}.
Ot{\'{\i}} Floranes et al.\,\citep{oti2005} computed kernels for these
quantities and demonstrated that they are, indeed, largely insensitive
to the near-surface layers and they have seen widespread use in
asteroseismic modelling.  From the same underlying principles,
Roxburgh\,\citep{roxburgh2015, roxburgh2016} described methods to fit
out a more general $\ell$-independent component of the frequency
differences.  These are too new to have been used widely.

These various parametrized methods have not yet been systematically
compared with observations, though the community's collective
experience suggests that none generally leads to absurd results.
Schmitt \& Basu\,\citep{schmitt2015} conducted the most thorough study
yet by inserting structural perturbations into stellar models across
the HR diagram and then trying to fit the frequency differences using
the solar-calibrated power law, the cubic and combined terms of Ball
\& Gizon\,\citep{ball2014} or the observed solar surface effect,
rescaled by the large separation.  The combined term by Ball \&
Gizon\,\citep{ball2014} appeared to fare best, although the scaled
solar term also performed reasonably on the main sequence.

The parametrizations do not solve the problem of the surface effects
but they at least allow us to exploit the reams of data already
available while we work towards properly modelling the surface
effects.  The results should always be interpreted with the knowledge
that the Sun remains the only star for which we can truly calibrate
the frequency differences.  Everything else depends on the confidence
we place in how well our best-fitting models represent the stars under
study.

\begin{figure}
\centering
\includegraphics[width=\hsize,clip]{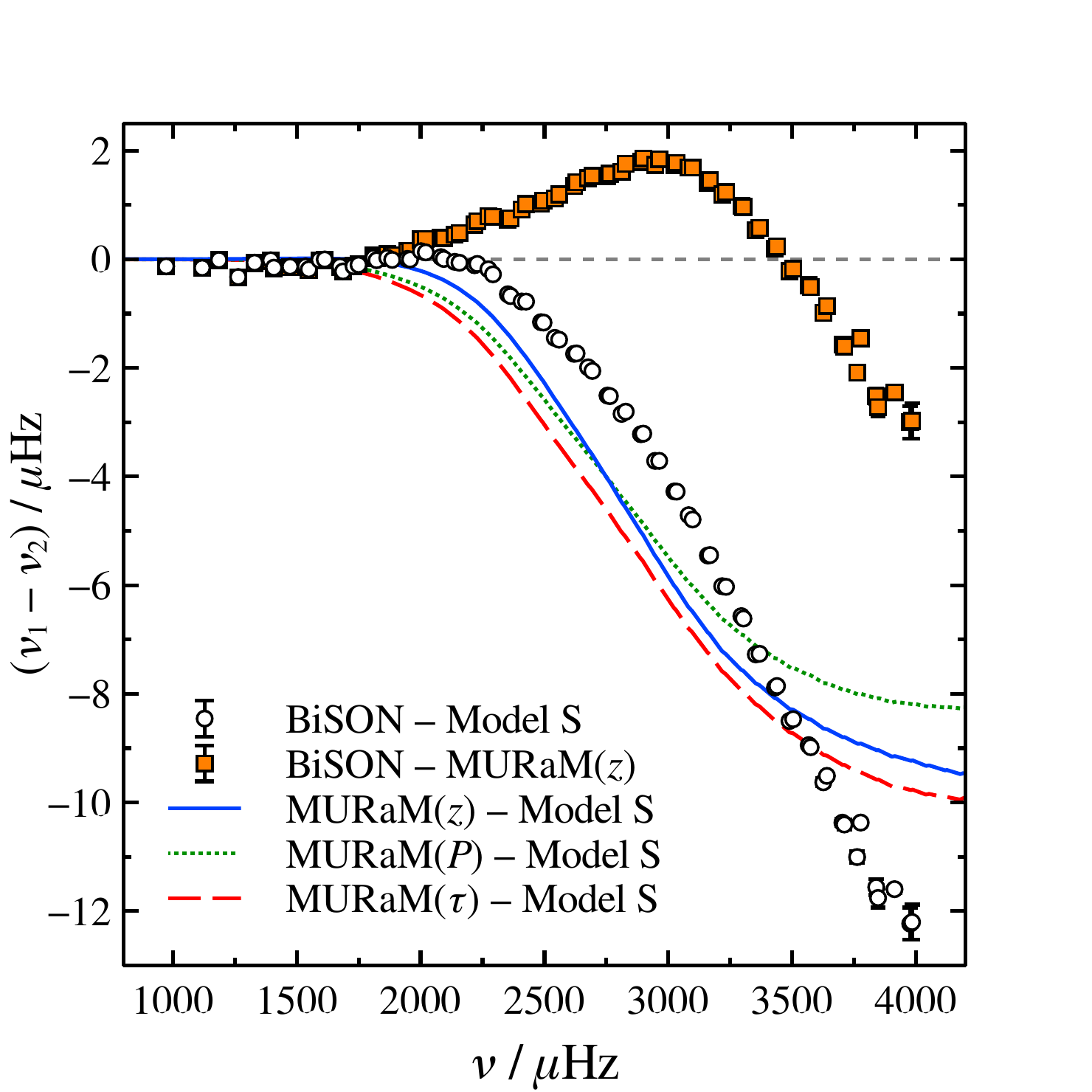}
\caption{Frequency differences for the Sun between different
  combinations of models and data as a function of model frequencies.
  The lines show the frequency differences between the solar model
  before and after the near-surface layers are replaced by
  horizontally-averaged 3D RHD simulation data.  The different lines
  correspond to different choices of averaging co-ordinate: geometric
  depth $z$ (solid blue), pressure $P$ (dotted green) or optical depth
  $\tau$ (dashed red).  The white points show the same differences
  between observed and modelled frequencies as in
  Fig.~\ref{f:problem}.  The orange squares show the frequency
  differences after the near-surface layers of the solar model have
  been replaced by 3D RHD simulation data averaged at constant
  geometric depth $z$.  The overall extent of the surface effect is
  reduced to a few $\mu\mathrm{Hz}$ but a clear systematic difference
  remains.}
\label{f:3drhd}
\end{figure}

\begin{figure}
\centering
\includegraphics[width=\hsize,clip]{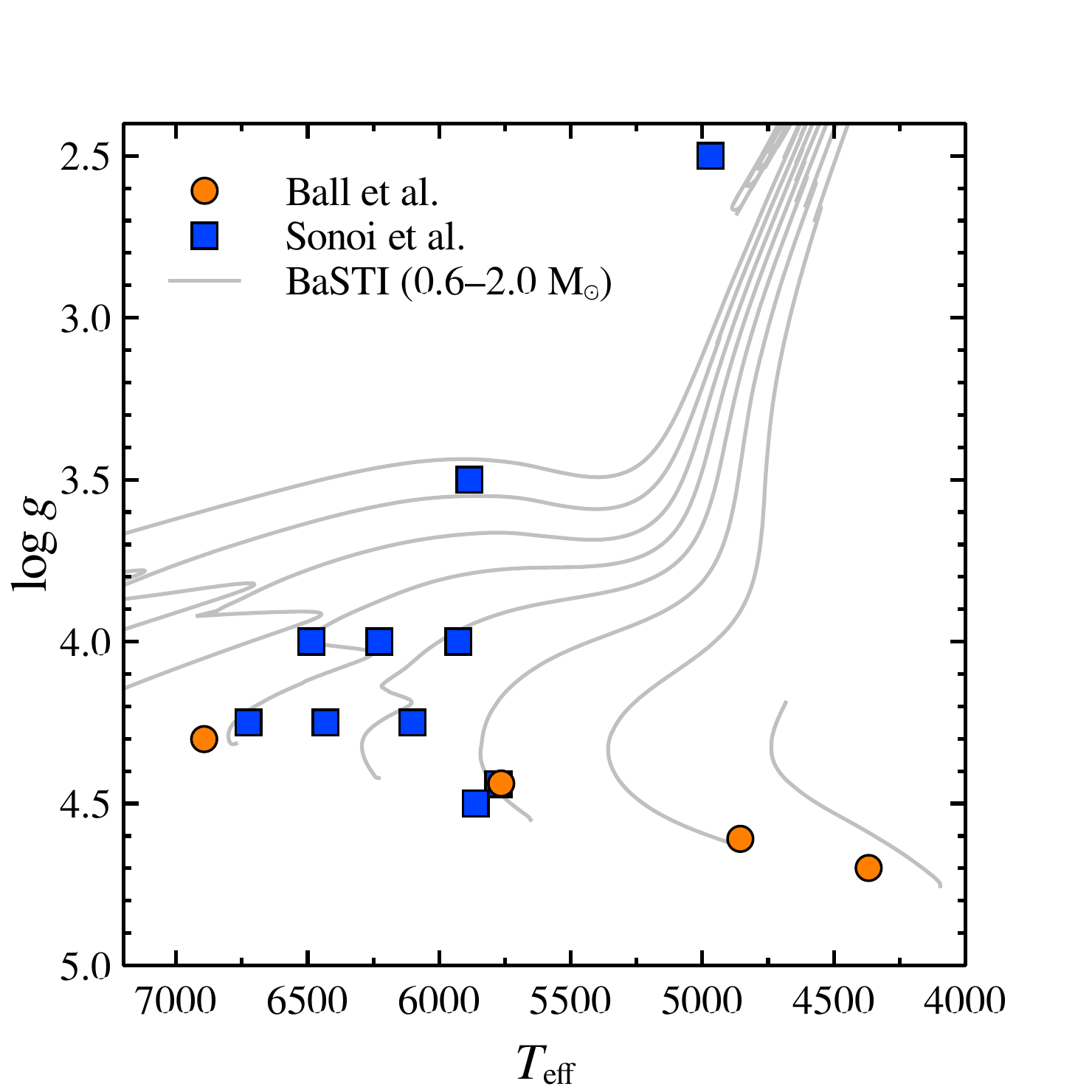}
\caption{Kiel diagram (surface gravity against effective temperature)
  showing the parameters of the 3D RHD simulations used by \citep[blue
  squares]{sonoi2015} and \citep[orange circles]{ball2016}.  The grey
  lines show solar-metallicity evolutionary tracks from BaSTI
  \citep{basti2004} for masses from $0.6$ to $2.0\Msun$ in steps of
  $0.2\Msun$.  Both sets of simulations include a solar model (models
  A and G2 in Sonoi et al.\,\citep{sonoi2015} and Ball et al.\,\citep{ball2016}) and the hottest
  models (models B and F3 in Sonoi et al.\,\citep{sonoi2015} and Ball et al.\,\citep{ball2016})
  also have comparable parameters.}
\label{f:HR}
\end{figure}

\begin{figure*}
\centering
\begin{tabular}{r@{\kern-3em}l}
\includegraphics[width=0.5\hsize]{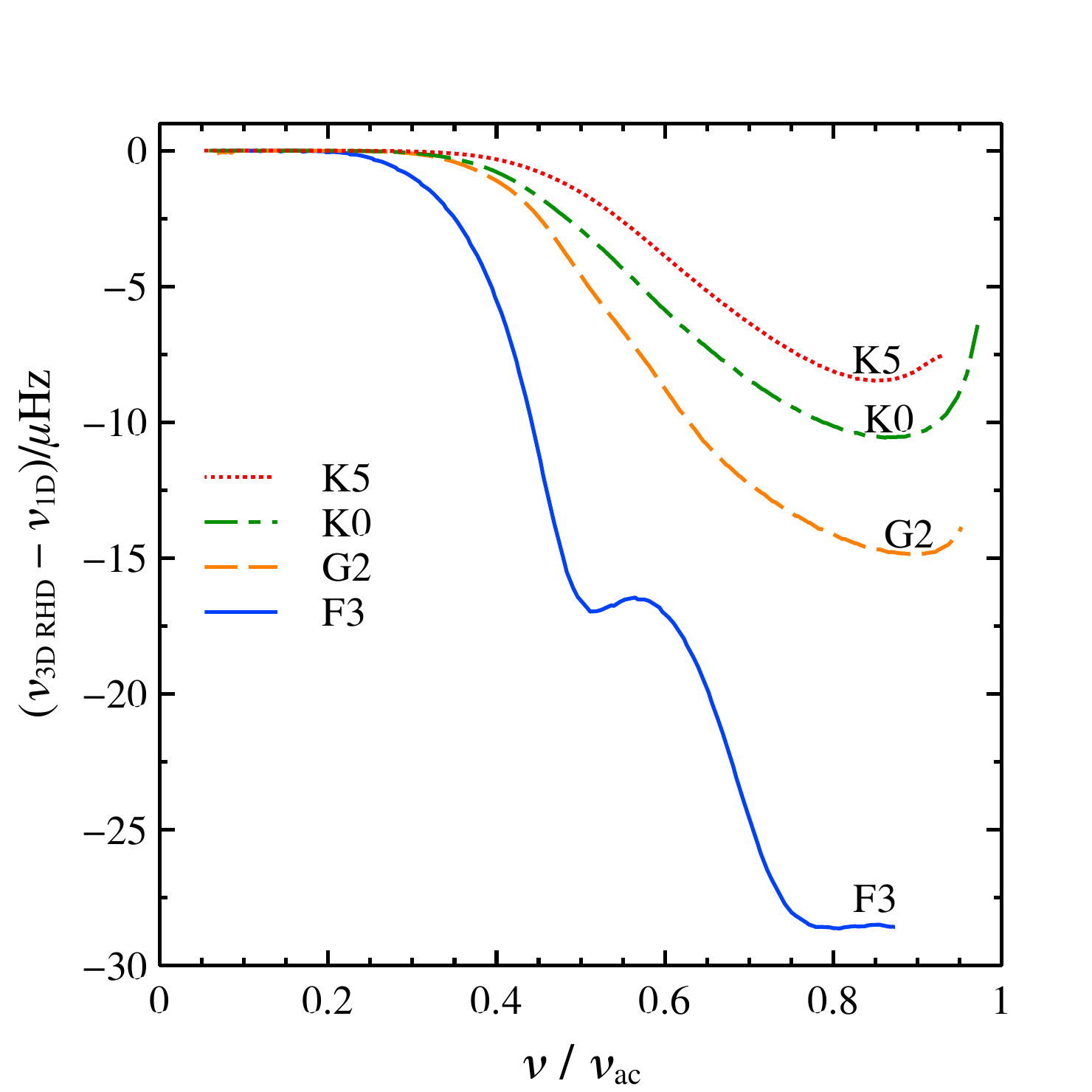} &
\includegraphics[width=0.5\hsize]{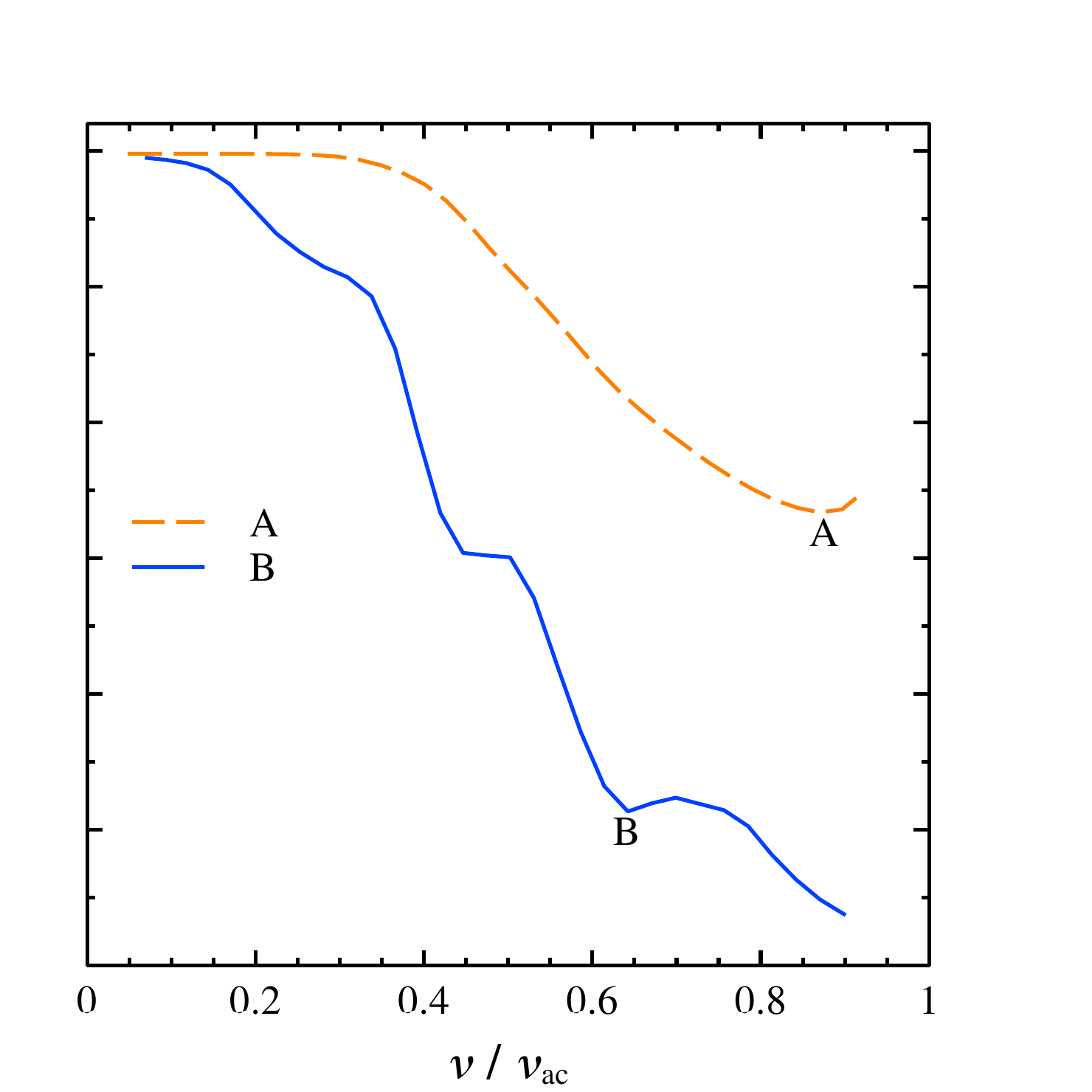}
\end{tabular}
\caption{Frequency differences between various models before and after
  their near-surface layers are replaced with horizontally-averaged 3D
  RHD simulation profiles, as a function of frequency normalized to
  the acoustic cut-off frequency.  The left panel shows models from
  Ball et al.\,\citep{ball2016}; the right panel for models from Sonoi
  et al.\,\citep[adapted from their Fig.~3]{sonoi2015}.  The models
  labelled A and B on the right correspond to the solar model and
  hottest model of Sonoi et al.\,\citep{sonoi2015} and can be compared
  to models G2 and F3 on the left, respectively.}
\label{f:spectype}
\end{figure*}

\section{Three-dimensional radiation hydrodynamics}
\label{s:3drhd}


I mentioned in Sec.~\ref{s:problem} that the surface effect is chiefly
caused by improper modelling of near-surface convection.  So why not
use better models of near-surface convection?  This is the idea behind
recent efforts to combine stellar models with 3D RHD simulations.
Several groups have simulated near-surface convection from first
principles in stars of various spectral types
\citep[e.g.][]{ludwig2009, beeck2013a, magic2013}.  These simulations
are sufficiently realistic to reproduce most of the observed
characteristics of the Sun's near-surface convection \citep[again, see
][for a review]{nordlund2009} and it is assumed that they are
similarly realistic for other stars.

The process of replacing a stellar model's near-surface layers with
averaged simulation data is becoming known as \emph{patching}.  The
frequency differences are then computed between the \emph{unpatched}
model (the original stellar model) and the \emph{patched} model (with
the near-surface layers replaced).

The idea of patching is not new.  Rosenthal et
al.\,\citep{rosenthal1999} restricted their study of solar
oscillations to modes with angular degree $\ell>60$.  These modes are
trapped within the solar convection zone, so they could compare their
averaged simulation data with envelope models computed using MLT.
Their early results showed that replacing the equilibrium structure of
the stellar model with the simulation data, averaged at constant
geometric depth, already introduced a surface effect of similar
magnitude to the observed effect, although a significant systematic
effect remained.  More recently, Piau et al.\,\citep{piau2014} used 3D
RHD simulation data to compute surface effects in a complete solar
model (not just the convective envelope), finding that the structural
component of the surface effect reduced the remain discrepancy to a
few $\uHz$.  Finally, on the subject of the solar surface correction,
Magic \& Weiss\,\citep{magic2016} computed surface effects using
simulations with different input magnetic field strengths and found
that they could reproduce reasonably well the frequency shifts induced
by the changing level of magnetic activity in the Sun.


The 12 months preceding this meeting saw the first papers to combine
stellar models and 3D RHD simulations for the surface effects in other
types of star.  First, Sonoi et al.\,\citep{sonoi2015} combined stellar models from
CESTAM \citep{morel1997,morel2008} with simulations from the CIFIST
atmosphere grid \citep{ludwig2009}.  Second, Ball et al.\,\citep{ball2016} combined
stellar models from MESA \citep{paxton2011,paxton2013,paxton2015} with
simulations from the MURaM code \citep{beeck2013a}.  The two groups
independently performed nearly the same calculations using somewhat
complementary sets of stellar models.  The ten simulations used by
Sonoi et al.\,\citep{sonoi2015} cover one red giant (around the red clump) and
dwarfs and subgiants hotter than the Sun.  The four simulations used
by Ball et al.\,\citep{ball2016} span the main-sequence from spectral type F3 to
K5.  Fig.~\ref{f:HR} shows the atmospheric parameters for the two
groups' simulations.  

These studies are not definitive.  For a start, they only deal with
the part of the surface effect caused by improving the structure of
the equilibrium stellar model.  The averaged simulation profiles
include the turbulent pressure but it remains unclear what is the
appropriate form of the perturbation to the turbulent pressure.  Both
Sonoi et al.\,\citep{sonoi2015} and Ball et al.\,\citep{ball2016}
assume that the turbulent pressure varies with the total pressure:
Rosenthal et al.\,\citep{rosenthal1999} dubbed this the \emph{gas
  gamma one} approximation.  Changing this assumption potentially
affects the results by a factor of about two \citep{rosenthal1999}.
Moreover, it is unclear exactly what is the appropriate horizontal
average to take from the simulation data.  Sonoi et
al.\,\citep{sonoi2015} and Ball et al.\,\citep{ball2016} both used
averages over constant geometric depth but different averages give
surface effects that differ by a few $\uHz$ for the Sun (see
Fig.~\ref{f:3drhd}).

Fig.~\ref{f:3drhd} shows how the mode frequencies of a standard solar
model \citep[Model S again,][]{modelS} are changed by the modification
of the near-surface equilibrium structure.  The white points are the
same differences shown in Figs~\ref{f:problem} and \ref{f:param}.  The
solid blue, dotted green and dashed red curves are the differences in
the model frequencies before and after patching with the G2-type MURaM
simulation averaged over constant geometric depth, pressure or optical
depth.  The spread in the curves shows that uncertainty above the
appropriate average introduces an uncertainty in the frequency shifts
of about $0.5$--$1.0\uHz$.  Finally, the orange squares show the
remaining difference between the patched model (with the simulation
averaged at constant geometric depth) and the BiSON observations.  The
overall surface effect is reduced substantially, though clearly a
large effect remains and the remaining difference is still a surface
effect.  It is not yet clear if the remaining trend is because the
averaged near-surface structure is still not quite right, because
non-adiabatic effects have been neglected, or (most likely) both.

With these uncertainties in mind, both teams found that the surface
effect is larger in stars that are hotter.  Based on their cooler
dwarfs, Ball et al.\,\citep{ball2016} also noted that the overall
shape of the frequency differences as a function of frequency is
similar in the G2-, K0- and K5-type models, but some qualitative
change sets in between the F3- and G2-type models.  Sonoi et
al.\,\citep{sonoi2015}, with their greater coverage of surface
gravity, also found that the surface effect increases with increasing
surface gravity.  Within their limitations, the two groups' results
are mutually consistent.  They have two simulations with similar
parameters and the results agree well.  Fig.~\ref{f:spectype} shows
the frequency shifts for all the simulations by Ball et
al.\,\citep[][left]{ball2016} and the simulations A and B of Sonoi et
al.\,\citep[][right]{sonoi2015}, which have similar parameters to
models G2 and F3 of Ball et al.\,\citep{ball2016}.  

Both teams also compared the parametric fits described in
Sec.~\ref{s:param}, though Sonoi et al.\,\citep{sonoi2015} only
compared their modified Lorentzian with a power law.
Fig.~\ref{f:spectype} shows that a simple power law does not describe
the differences between the patched and unpatched models very well.
Ball et al.\,\citep{ball2016} found that the combined term was
consistently superior, notably in their F3 model, but a scaled solar
frequency correction fits reasonably well in their three cooler
models.  Sonoi et al.\,\citep{sonoi2015} provided simple fits to the
best-fit parameters as a function of surface properties, though they
did not consider a scaled solar correction or either of the
corrections by Ball \& Gizon \citep{ball2014}.

Further exploitation of the 3D RHD simulations is underway, notably on
non-adiabatic effects, but these early results already give some
indication of how much of a surface effect is introduced by improving
the background stellar model.  It remains to be seen if the
conclusions hold up as further surface effects are considered.

\section{The future}
\label{s:future}

To close, I briefly opine on how we might progress further on the
problem of surface effects.  The main theoretical path at this point
is to further exploit the 3D RHD simulations.  There is far more
information available than simply the horizontally- and
temporally-averaged profiles and this information can be used to
investigate other components of the surface effect.  But it should be
remembered that even indirect conclusions drawn from the simulations
can be useful.  For example, the parametrizations of the surface
effects tend to correlate with the mixing-length parameter in stellar
models.  There is good physical reason for this: both the surface
effect and the mixing-length parameter are sensitive to the
superadiabatic layer near the stellar surface.  If the mixing-length
parameter is constrained separately by the simulations
\citep[e.g.][]{ludwig1999, trampedach2014b} then the surface effect is
also better constrained.

Progress is more difficult from the observational side.  The best
solar-like oscillators from the nominal \emph{Kepler} mission show
modes oscillating at frequencies nearly low enough that they are
unaffected by the surface effect.  If just a few more radial orders
could be detected, these low frequencies could potentially be used to
fit models without a surface term, though at the cost of discarding
the many higher-frequency modes that are available.  Alas, no imminent
mission will provide such high-quality data for single targets, so we
may have to wait until PLATO \citep{plato} for higher-quality data on
single targets.

From the ground, however, there is tremendous potential from the
Stellar Oscillation Network Group \citep[see e.g.][these
proceedings]{jcd_kasc9}.  Because it observes in radial velocity, the
background signal of granulation is weaker, which allows
lower-frequency modes to be detected more easily.  This could allow us
to calibrate models directly to the unaffected frequencies and inspect
the remaining frequencies to determine the surface effect after
fitting the stellar model.  One node of the network is fully
operational and another partially so.  The first results from the
first node were reported at this meeting \citep{jcd_kasc9}.  Adding
nodes to the network probably represents our best chance of bringing
tight observational constraints to bear on the problem of surface
effects.

{\small
  \section*{Acknowledgement}
  
  The author would like to thank the organizers for partial financial
  support to attend the meeting.  He also acknowledges research
  funding by Deutsche Forschungsgemeinschaft (DFG) under grant SFB
  963/1 ``Astrophysical flow instabilities and turbulence'', Projects
  A18.}

\bibliography{../master}

\begin{thebibliography}{40}

\bibitem{corot}
M.~{Auvergne}, P.~{Bodin}, L.~{Boisnard}, J.T. {Buey}, S.~{Chaintreuil},
  G.~{Epstein}, M.~{Jouret}, T.~{Lam-Trong}, P.~{Levacher}, A.~{Magnan} et~al.,
  A\&A, \textbf{506}, 411 (2009), \texttt{0901.2206}

\bibitem{kepler}
W.J. {Borucki}, D.~{Koch}, G.~{Basri}, N.~{Batalha}, T.~{Brown}, D.~{Caldwell},
  J.~{Caldwell}, J.~{Christensen-Dalsgaard}, W.D. {Cochran}, E.~{DeVore}
  et~al., Science \textbf{327}, 977 (2010)

\bibitem{broomhall2009}
A.M. {Broomhall}, W.J. {Chaplin}, G.R. {Davies}, Y.~{Elsworth}, S.T.
  {Fletcher}, S.J. {Hale}, B.~{Miller}, R.~{New}, MNRAS, \textbf{396}, L100
  (2009), \texttt{0903.5219}

\bibitem{davies2014a}
G.R. {Davies}, W.J. {Chaplin}, Y.~{Elsworth}, S.J. {Hale}, MNRAS, \textbf{441},
  3009 (2014), \texttt{1405.0160}

\bibitem{modelS}
J.~{Christensen-Dalsgaard}, W.~{Dappen}, S.V. {Ajukov}, E.R. {Anderson}, H.M.
  {Antia}, S.~{Basu}, V.A. {Baturin}, G.~{Berthomieu}, B.~{Chaboyer}, S.M.
  {Chitre} et~al., Science \textbf{272}, 1286 (1996)

\bibitem{kjeldsen1995}
H.~{Kjeldsen}, T.R. {Bedding}, A\&A \textbf{293}, 87 (1995),
  \texttt{arXiv:astro-ph/9403015}

\bibitem{nordlund2009}
{\AA}.~{Nordlund}, R.F. {Stein}, M.~{Asplund}, Living Reviews in Solar Physics
  \textbf{6} (2009)

\bibitem{rosenthal1997}
C.S. {Rosenthal}, \emph{{Convective Effects on Mode Frequencies}}, in
  \emph{SCORe'96 : Solar Convection and Oscillations and their Relationship},
  edited by F.P. {Pijpers}, J.~{Christensen-Dalsgaard}, C.S. {Rosenthal}
  (1997), Vol. 225 of \emph{Astrophysics and Space Science Library}, pp.
  145--160

\bibitem{rosenthal1999}
C.S. {Rosenthal}, J.~{Christensen-Dalsgaard}, {\AA}.~{Nordlund}, R.F. {Stein},
  R.~{Trampedach}, A\&A, \textbf{351}, 689 (1999), \texttt{astro-ph/9803206}

\bibitem{brown1984}
T.M. {Brown}, Science \textbf{226}, 687 (1984)

\bibitem{houdek1996}
G.~{Houdek}, Ph.D.~Thesis, Formal- und Naturwisseschaftliche
  Fakult{\"a}t der Universit{\"a}t Wien, (1996) (1996)

\bibitem{kbcd2008}
H.~{Kjeldsen}, T.R. {Bedding}, J.~{Christensen-Dalsgaard}, ApJL, \textbf{683},
  L175 (2008), \texttt{0807.1769}

\bibitem{ball2014}
W.H. {Ball}, L.~{Gizon}, A\&A, \textbf{568}, A123 (2014), \texttt{1408.0986}

\bibitem{sonoi2015}
T.~{Sonoi}, R.~{Samadi}, K.~{Belkacem}, H.G. {Ludwig}, E.~{Caffau},
  B.~{Mosser}, A\&A, \textbf{583}, A112 (2015), \texttt{1510.00300}

\bibitem{silva2013}
V.~{Silva Aguirre}, S.~{Basu}, I.M. {Brand{\~a}o}, J.~{Christensen-Dalsgaard},
  S.~{Deheuvels}, G.~{Do{\u g}an}, T.S. {Metcalfe}, A.M. {Serenelli},
  J.~{Ballot}, W.J. {Chaplin} et~al., ApJ, \textbf{769}, 141 (2013),
  \texttt{1304.2772}

\bibitem{ledoux1958}
P.~{Ledoux}, T.~{Walraven}, Handbuch der Physik \textbf{51}, 353 (1958)

\bibitem{gough1990}
D.O. {Gough}, \emph{{Comments on Helioseismic Inference}}, in \emph{Progress of
  Seismology of the Sun and Stars}, edited by Y.~{Osaki}, H.~{Shibahashi}
  (Springer Verlag, Berlin, 1990), Vol. 367 of \emph{Lecture Notes in Physics},
  p. 283

\bibitem{libbrecht1990}
K.G. {Libbrecht}, M.F. {Woodard}, Nature, \textbf{345}, 779 (1990)

\bibitem{goldreich1991}
P.~{Goldreich}, N.~{Murray}, G.~{Willette}, P.~{Kumar}, ApJ, \textbf{370}, 752
  (1991)

\bibitem{roxburgh2003}
I.W. {Roxburgh}, S.V. {Vorontsov}, A\&A, \textbf{411}, 215 (2003)

\bibitem{roxburgh2015}
I.W. {Roxburgh}, A\&A, \textbf{574}, A45 (2015), \texttt{1406.6491}

\bibitem{roxburgh2016}
I.W. {Roxburgh}, A\&A, \textbf{585}, A63 (2016)

\bibitem{oti2005}
H.~{Ot{\'{\i}} Floranes}, J.~{Christensen-Dalsgaard}, M.J. {Thompson}, MNRAS,
  \textbf{356}, 671 (2005)

\bibitem{schmitt2015}
J.R. {Schmitt}, S.~{Basu}, ApJ, \textbf{808}, 123 (2015), \texttt{1506.06678}

\bibitem{ball2016}
W.H. {Ball}, B.~{Beeck}, R.H. {Cameron}, L.~{Gizon}, A\&A, \textbf{592}, A159
  (2016), \texttt{1606.02713}

\bibitem{basti2004}
A.~{Pietrinferni}, S.~{Cassisi}, M.~{Salaris}, F.~{Castelli}, ApJ,
  \textbf{612}, 168 (2004), \texttt{astro-ph/0405193}

\bibitem{ludwig2009}
H.G. {Ludwig}, E.~{Caffau}, M.~{Steffen}, B.~{Freytag}, P.~{Bonifacio},
  A.~{Ku{\v c}inskas}, Mem. Soc. Astron. Italiana, \textbf{80}, 711 (2009), \texttt{0908.4496}

\bibitem{beeck2013a}
B.~{Beeck}, R.H. {Cameron}, A.~{Reiners}, M.~{Sch{\"u}ssler}, A\&A,
  \textbf{558}, A48 (2013), \texttt{1308.4874}

\bibitem{magic2013}
Z.~{Magic}, R.~{Collet}, M.~{Asplund}, R.~{Trampedach}, W.~{Hayek},
  A.~{Chiavassa}, R.F. {Stein}, {\AA}.~{Nordlund}, A\&A, \textbf{557}, A26
  (2013), \texttt{1302.2621}

\bibitem{piau2014}
L.~{Piau}, R.~{Collet}, R.F. {Stein}, R.~{Trampedach}, P.~{Morel},
  S.~{Turck-Chi{\`e}ze}, MNRAS, \textbf{437}, 164 (2014), \texttt{1309.7179}

\bibitem{magic2016}
Z.~{Magic}, A.~{Weiss}, A\&A, \textbf{592}, A24 (2016), \texttt{1606.01030}

\bibitem{morel1997}
P.~{Morel}, A\&AS, \textbf{124} (1997)

\bibitem{morel2008}
P.~{Morel}, Y.~{Lebreton}, Ap\&SS, \textbf{316}, 61 (2008), \texttt{0801.2019}

\bibitem{paxton2011}
B.~{Paxton}, L.~{Bildsten}, A.~{Dotter}, F.~{Herwig}, P.~{Lesaffre},
  F.~{Timmes}, ApJS, \textbf{192}, 3 (2011), \texttt{1009.1622}

\bibitem{paxton2013}
B.~{Paxton}, M.~{Cantiello}, P.~{Arras}, L.~{Bildsten}, E.F. {Brown},
  A.~{Dotter}, C.~{Mankovich}, M.H. {Montgomery}, D.~{Stello}, F.X. {Timmes}
  et~al., ApJS, \textbf{208}, 4 (2013), \texttt{1301.0319}

\bibitem{paxton2015}
B.~{Paxton}, P.~{Marchant}, J.~{Schwab}, E.B. {Bauer}, L.~{Bildsten},
  M.~{Cantiello}, L.~{Dessart}, R.~{Farmer}, H.~{Hu}, N.~{Langer} et~al., ApJS,
  \textbf{220}, 15 (2015), \texttt{1506.03146}

\bibitem{ludwig1999}
H.G. {Ludwig}, B.~{Freytag}, M.~{Steffen}, A\&A, \textbf{346}, 111 (1999),
  \texttt{astro-ph/9811179}

\bibitem{trampedach2014b}
R.~{Trampedach}, R.F. {Stein}, J.~{Christensen-Dalsgaard}, {\AA}.~{Nordlund},
  M.~{Asplund}, MNRAS, \textbf{445}, 4366 (2014), \texttt{1410.1559}

\bibitem{plato}
H.~{Rauer}, C.~{Catala}, C.~{Aerts}, T.~{Appourchaux}, W.~{Benz},
  A.~{Brandeker}, J.~{Christensen-Dalsgaard}, M.~{Deleuil}, L.~{Gizon}, M.J.
  {Goupil} et~al., Experimental Astronomy \textbf{38}, 249 (2014),
  \texttt{1310.0696}

\bibitem{jcd_kasc9}
J.~{Christensen-Dalsgaard}, \emph{200 nights with $\mu$ Herculis: early results
  from the SONG Hertzsprung telescope}, in \emph{Seismology of the Sun and the
  Distant Stars 2016}, edited by J.P.F.G. {Monteiro}, M.S. {Cunha}, J.M.T.
  {Ferreira} (2017), this volume

\end{thebibliography}

\end{document}